\documentclass[prl,aps,floats,superscriptaddress,floatfix,twocolumn]{revtex4}
\usepackage{amssymb,amsmath}
\usepackage{amsmath,amssymb}
\usepackage{graphicx}
\usepackage{psfrag}
\usepackage{color}
\usepackage{soul}
\usepackage{dcolumn}
\usepackage{mathtools}

\usepackage{bm}
\usepackage[normalem]{ulem}

\def\beq{\begin{equation}}
\def\eeq{\end{equation}}
\def\bea{\begin{eqnarray}}
\def\eea{\end{eqnarray}}
\begin{document}
\title{
Correlated disorder versus correlated noise: Ordering in active systems}
 \author{Sudip Mukherjee}\email{sudip.bat@gmail.com}
\affiliation{Barasat Government College,
10, KNC Road, Gupta Colony, Barasat, Kolkata 700124,
West Bengal, India}
\author{Abhik Basu}\email{abhik.123@gmail.com,abhik.basu@saha.ac.in}
\affiliation{Theoretical Physics Division, Saha Institute of
Nuclear Physics, 1/AF Bidhannagar, Calcutta 700064, West Bengal, India}

\begin{abstract}
Can quenched disorder generate ordering in driven systems? Using a recently proposed hydrodynamic model, we show that sufficiently long-ranged quenched disorder can induce long-range order in two-dimensional (2D) nonreciprocal XY systems, even when the clean system exhibits only short-range order at finite noise. Active surfaces tangentially advected by quenched velocities, governed by the same hydrodynamic equation, become statistically flat with super- or subdiffusive relaxation. In three dimensions (3D), quenched disorder combined with nonreciprocity produces a novel transition between strong-coupling and asymptotically noninteracting regimes, supporting either long- or short-range order. In both 2D and 3D, the exponents are nonuniversal, which vary continuously with the degree of transversality of the quenched disorder. The degree of transversality of the quenched disorder can be tuned to induce transitions in the model for fixed disorder and noise variances.

\end{abstract}

\maketitle

Nonreciprocal interactions provide a generic route to nonequilibrium collective behavior, enabling dynamical phenomena with no equilibrium counterparts~\cite{nonrec1,nonrec2,nonrec3,nonrec4}. In particular, nonreciprocal variants of the two-dimensional (2D) XY model have attracted considerable interest as minimal descriptions of phase ordering and synchronization in systems where interactions between degrees of freedom are not constrained by action-reaction symmetry~\cite{klapp,sriram-defect,gambassi,solon,hanai,sm-ab}. Active surfaces provide another paradigmatic class of 2D nonequilibrium systems. These are thin, deformable interfaces that continuously consume energy to sustain driven states~\cite{bassereau-book}. They exhibit novel behaviors, such as instabilities and super-stiffness~\cite{tirtha-njp}, that have no equilibrium counterparts~\cite{nelson-book}.


Long-range noises in driven systems can drastically alter scaling behaviors with short-range noises. Canonical examples include stochastically driven models for fluid~\cite{kraichnan,yakhot,jkb-pramana,rahulrev} and magnetohydrodynamic~\cite{abmhd,mkv,abjkbmhd}, turbulent binary fluids~\cite{ab-binfluid} and related nonlinear models~\cite{sm-ab-multi} in comparison with fluids in thermal equilibrium~\cite{fns}. See also Ref.~\cite{medina} for effects of long-range noises on the Kardar-Parisi-Zhang (KPZ) equation~\cite{stanley} and Ref.~\cite{debayan-ckpz} for effects of long-range noises on conserved growth models~\cite{cates,sm-ab-mbe}. In another study, Ref.~\cite{feh}, spatially correlated annealed noise in a system of active particles can lead to formation of network patterns, not found without it. Such long-range noises can arise in a variety of ways, e.g., active agents with intrinsic magnetic~\cite{mag1,mag2} or electric~\cite{el1,el2} dipoles subject to fluctuating magnetic of electric fields having spatial correlations of desired structure, active colloids are exposed to a fluctuating light intensity field~\cite{col1,col2}. 

Quenched disorder provides yet another possible relevant effect on the scaling behavior in nonequilibrium systems. A recent well-known example involves demonstration of long range order (LRO) in 2D active flocks in the presence of quenched disorder~\cite{john-disorder1,john-disorder2,john-disorder3}. In another study of the quenched-disorder conserved KPZ equation, it was shown that the scaling of height fluctuations is affected by the scaling properties of the quenched disorder field~\cite{sm-ckpz}. In equilibrium systems, quenched randomness generally frustrates long-range order and can profoundly modify phase transitions. Its role in active and nonreciprocal systems, however, can be qualitatively different, as broken detailed balance allows disorder to couple to the relaxation of fluctuations in ways that are not possible in equilibrium systems. This raises a basic question: can scale-free quenched disorder suppress the order-destabilizing effects of long-range noise and thereby sustain LRO in driven systems?

\begin{figure}[htb]
 \includegraphics[width=0.48\columnwidth]{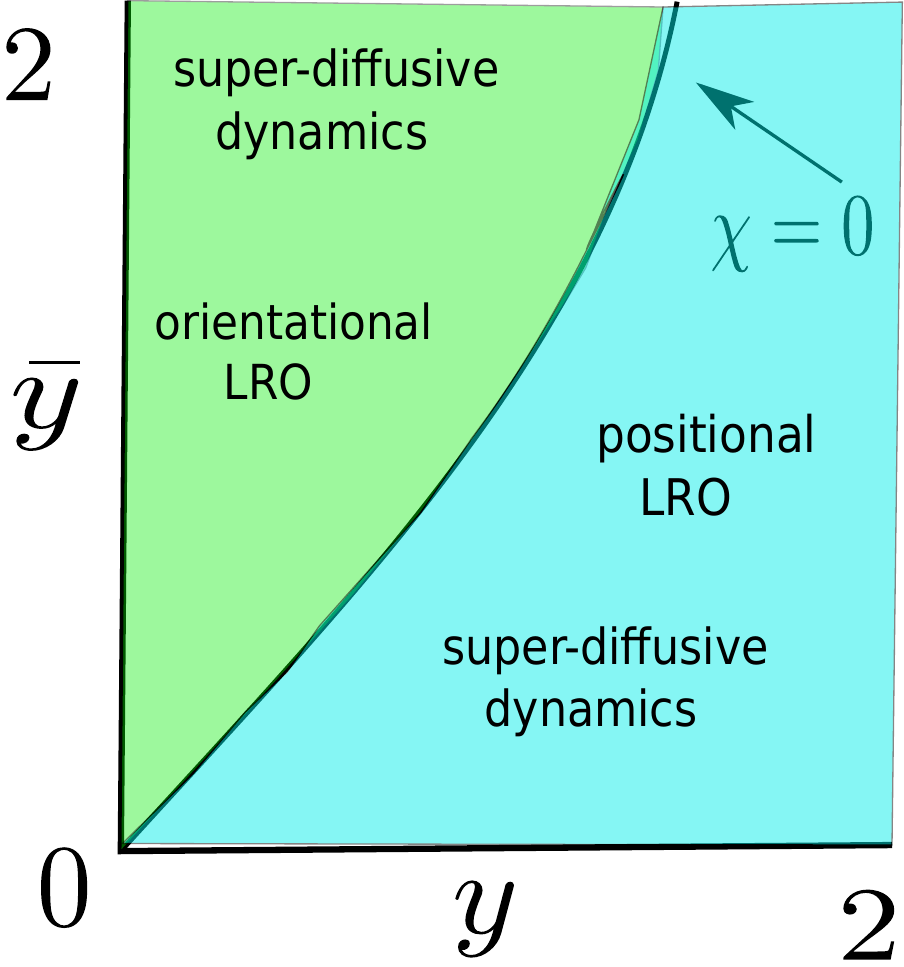}\hfill\includegraphics[width=0.49\columnwidth]{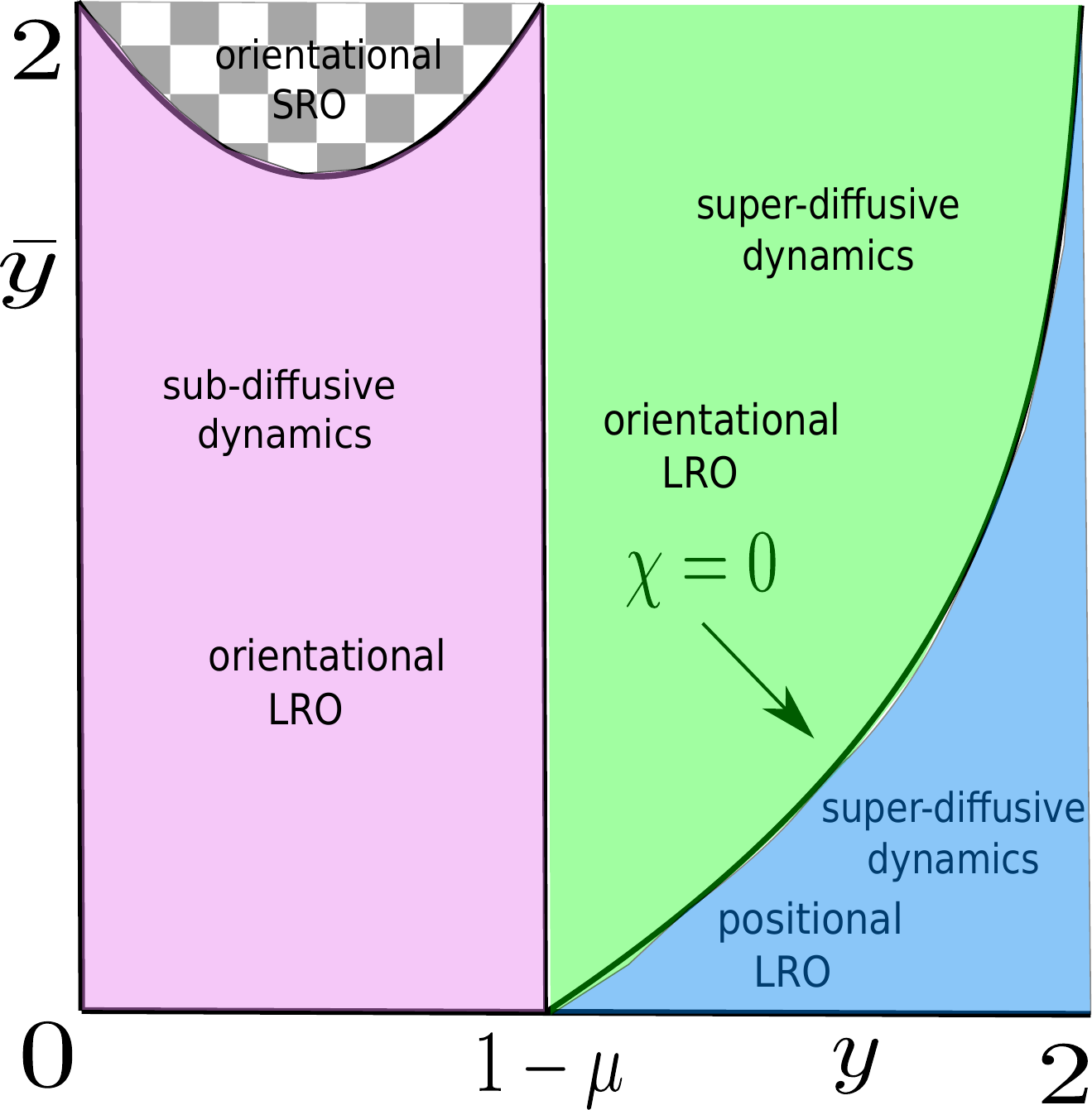}\\
 \includegraphics[width=0.49\columnwidth]{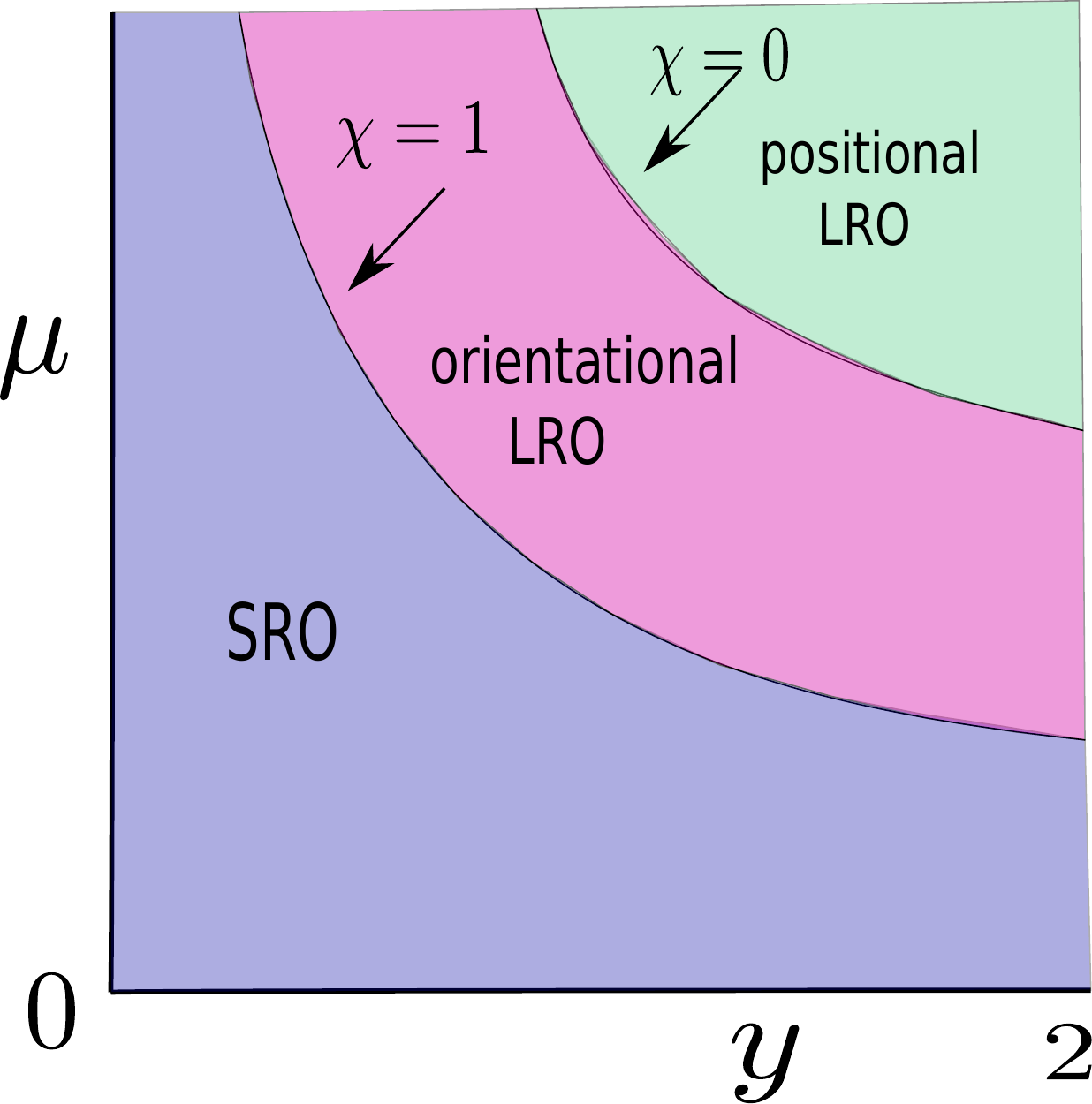}\hfill\includegraphics[width=0.49\columnwidth]{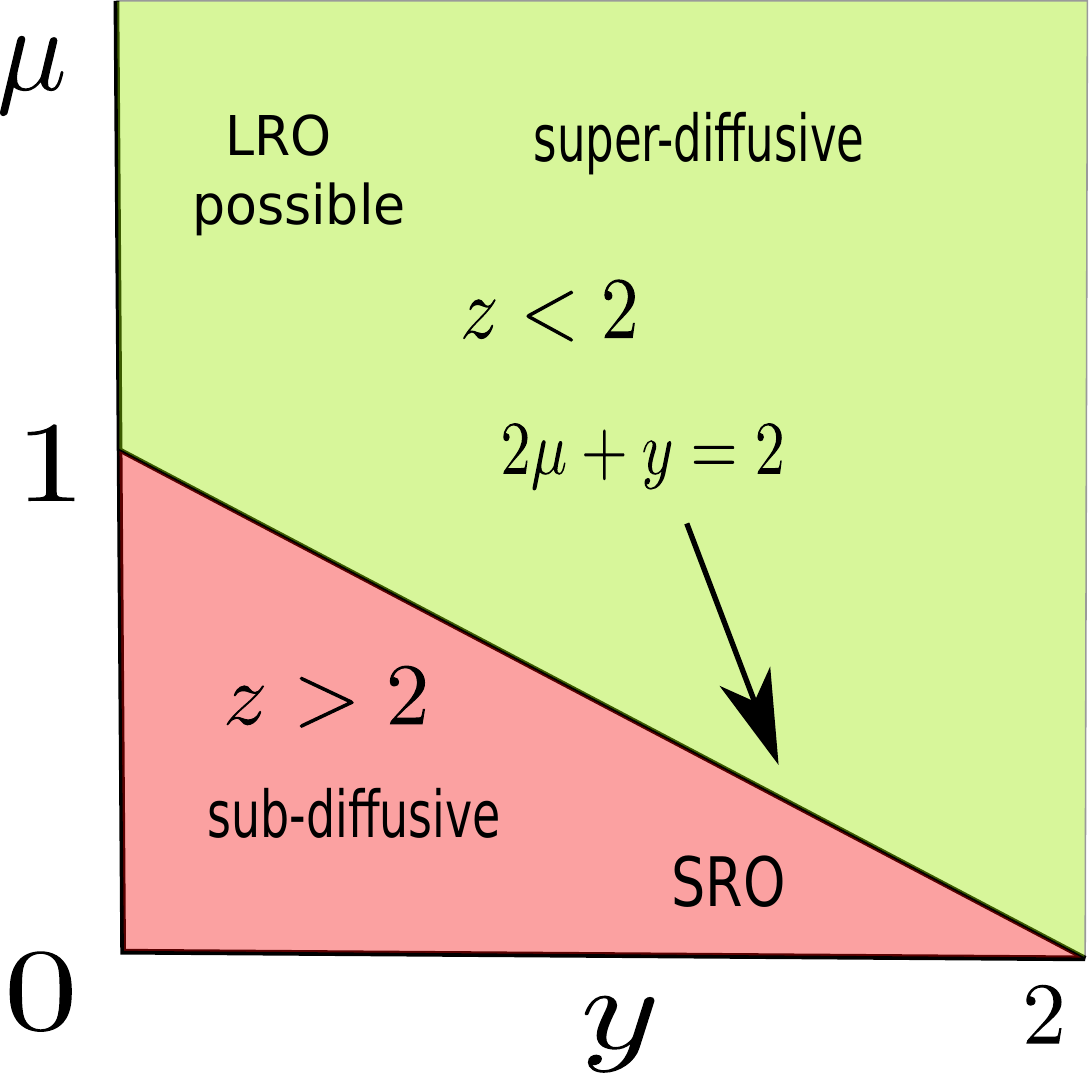}
 \caption{ Schematic phase diagrams in 2D for an active surface with quenched tangential velocity: (a) In the $y-\bar y$ plane for a fixed $\mu$ with $\mu>1$. Phases with positional LRO (blue) and orintational LRO but positional SRO (green) are marked. The phase boundary is given by $\chi=0$. The dynamics is super-diffusive everywhere. For the 2D NRB XY model, the phases are phase LRO (blue) and phase SRO (green). (b) For $y-\bar y$ plane for a fixed $\mu$ with $\mu<1$. Phases with positional LRO and super-diffusive dynamics (blue), orientational LRO but positional SRO with super-diffusive dynamics (green), orientational LRO but positional SRO  with sub-diffusive dynamics (red) and orientational SRO (checkerboard) are marked. The boundary between positional LRO  and orientaional LRO  are given by $\chi=0$, the boundary between orientaional SRO and orientaional LRO are given by $\chi=1$. For the 2D NRB XY model, the phases are LRO (green) and phase SRO (remaining phase diagram). (c) In the $y-\mu$ plane for a fixed for fixed $\bar y$. Phases with positional LRO (green), orientaional LRO, but positional SRO (magenta) and orientaional SRO (violate) for an active surface are marked. The phase boundary between positional LRO and orientational LRO, but positional SRO phases are given by $\chi=0$, and that between orientational LRO, but positional SRO and orientaional SRO phases is given by $\chi=1$. For the 2D NRB XY model, only phase LRO (green) and phase SRO (magenta and violet) exist. (d) Dynamics in the $y-\mu$ plane: yellow (super-diffusive, $z<2$) and red (sub-diffusive, $z>2$). The boundary between the two regions is given by $2\mu+y=2$. See text. }\label{phase2d}
\end{figure}
\begin{figure}[htb]
 \includegraphics[width=0.50\columnwidth]{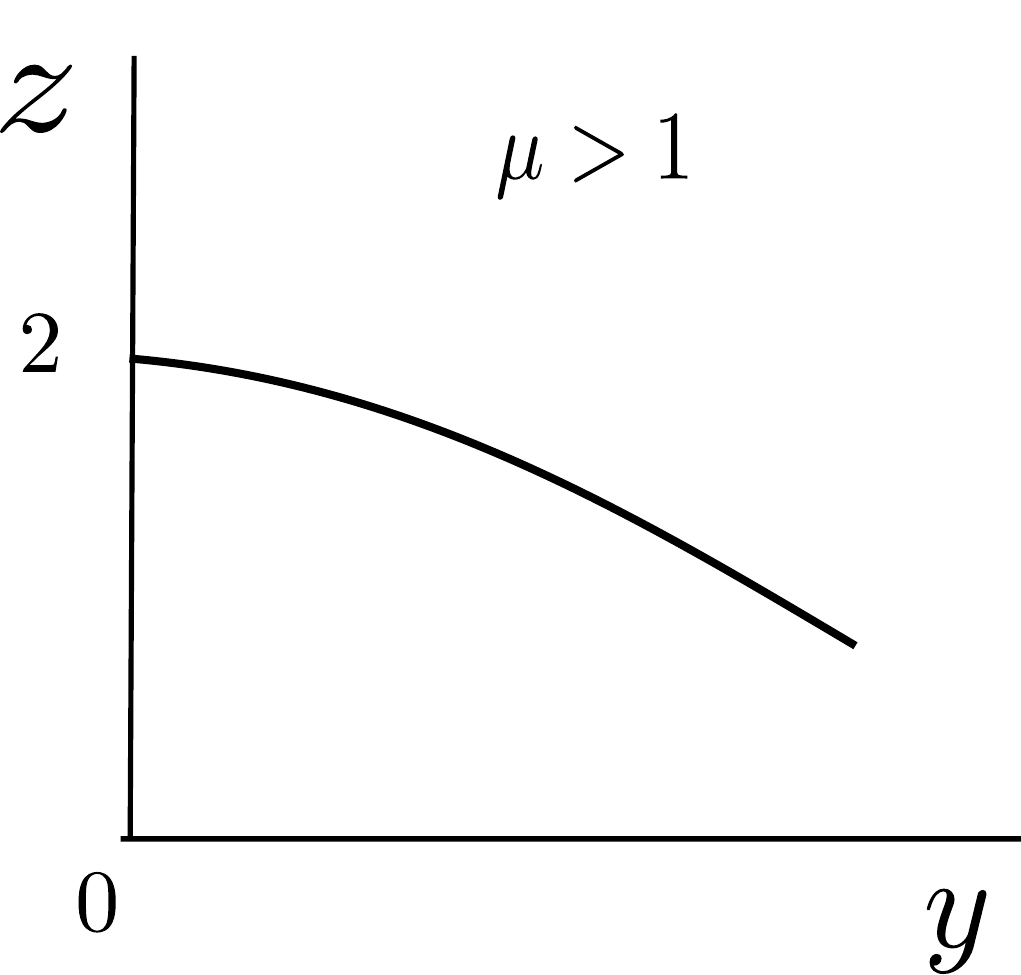}\hfill\includegraphics[width=0.50\columnwidth]{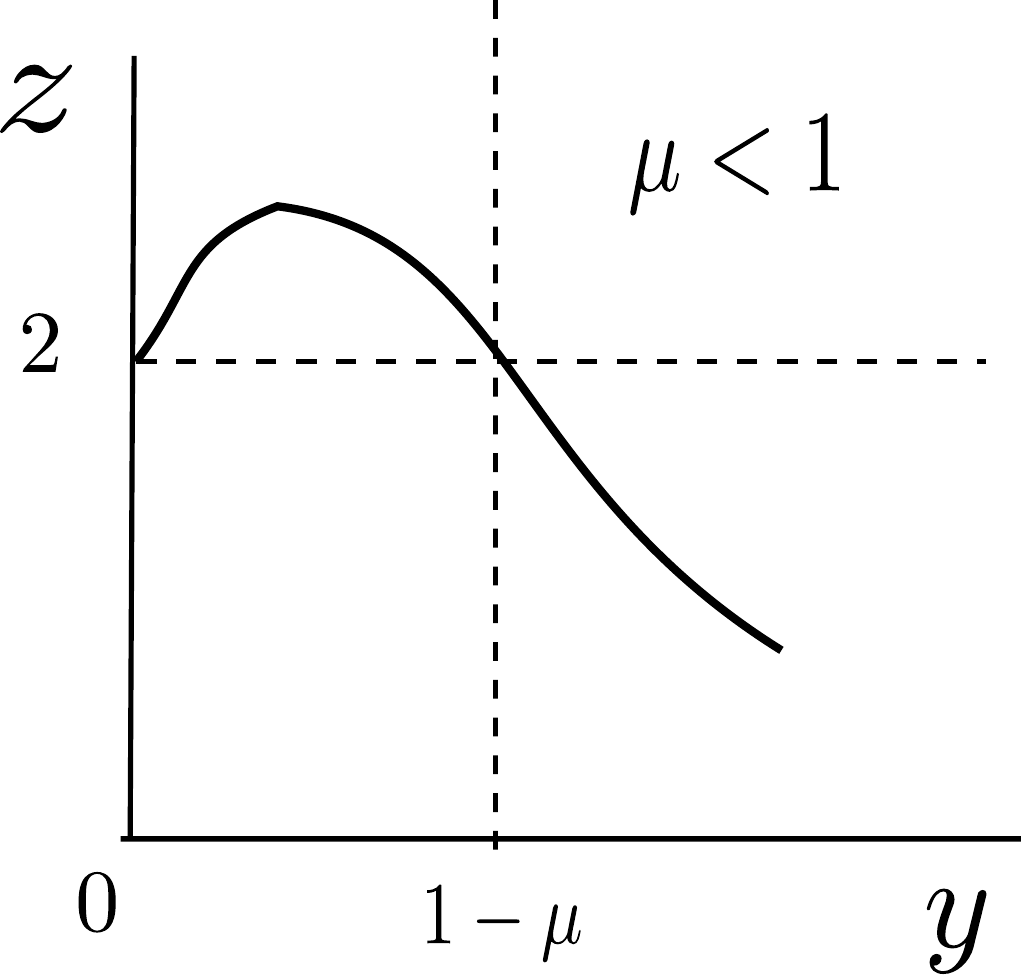}
 \caption{Variation of $z$ with $y$ in 2D for a fixed $\mu$. (a) $\mu>1$: $z$ monotonically decreases with $y$, starting from 2 at $y\rightarrow 0$. (b) $\mu<1$: $z$ first rises from 2 at $y\rightarrow 0$, then decreases monotonically with $y$, crossing 2 at $y=1-\mu$ to become less than 2. Thus $z$ has a non-monotonic dependence on $y$. See text.}\label{phase2d-z}
\end{figure}

In this Letter, we address this question by using the recently proposed common  hydrodynamic theory for a quenched disordered nonreciprocal random bond XY (hereafter NRB XY) model and an active inversion-symmetric surface advected by quenched tangential velocities~\cite{sm-ab}. The quenched disorder is assumed to be generated by a frozen heterogeneous embedding medium with the disorder being not independent at every point, but rather correlated over a distance, so that the local
asymmetry of bond-disordered interactions remains correlated over
distances much larger than the microscopic interaction
range. The dynamical degrees of freedom $\theta({\bf x},t)$, which is the phase in NRB XY model and height for the surface in Monge gauge~\cite{sm-ab}, are also subject to correlated environmental fluctuations. Such long-range spatially correlated noises are known to affect macroscopic properties of active systems; see, e.g., Ref.~\cite{gao}. Assume spatially scale-free disorder and noise variances, 
characterized by two exponents $y,\bar y>0$ respectively. 
%
Our central results, obtained using the hydrodynamic equation for $\theta({\bf x},t)$~\cite{sm-ab}, are: being controlled by $\mu$, a dimensionless parameter that characterize the quenched disorder distribution (see below for a more precise definition) sufficiently long-range quenched disorder, which are nonreciprocal bond disorder for the XY model and quenched tangential advecting velocity for a surface, can suppress the order destabilizing effects of long-range noise. Specifically in 2D,
(i) for $\overline y < \frac{y(2\mu-2+y)}{2(2\mu+y)}$, the NRB XY spins have orientational LRO and the active surface has positional LRO, with fast, super-diffusive relaxational dynamics. For $1>\bar y/2 - \frac{y(2\mu-2+y)}{2(2\mu+y)}>0$, the spin have short range order (SRO), but the surface has orientational LRO and positional SRO. This can occur with slow, sub-diffusive dynamics or fast, super-diffusive relaxational dynamics, which is controlled by $\mu$, a dimensionless ratio that determines the relative strength of the transverse and the compressive parts of the disorder variance (see below for a more precise definition). The above results are summarized in Fig.~\ref{phase2d}. Intriguingly, the dynamic exponent $z$ has a nonmonotonic dependence on $y$ for $\mu<1$; it first increases from its linear theory value of 2, then decreases continuously as $y$ inclreases. For $\mu>1$, $z$ decreases monotonically from $z=2$ as $y$ increases. The variation of the dynamic exponent $z$ with $y$ in 2D for $\mu>1$ and $\mu<1$ are shown in Fig.~\ref{phase2d-z}.

\begin{figure}[htb]
 \includegraphics[width=0.49\columnwidth]{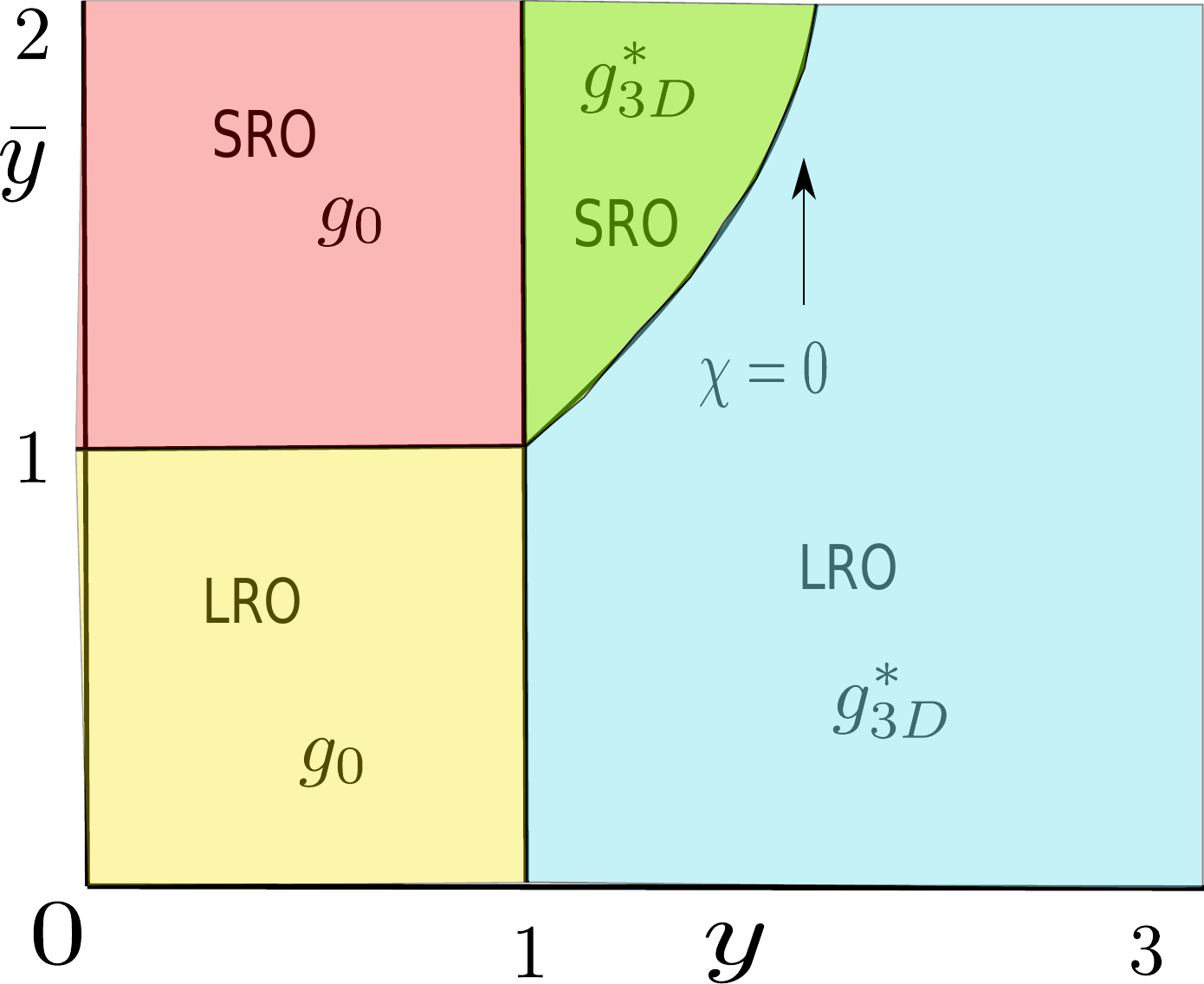}\hfill \includegraphics[width=0.51\columnwidth]{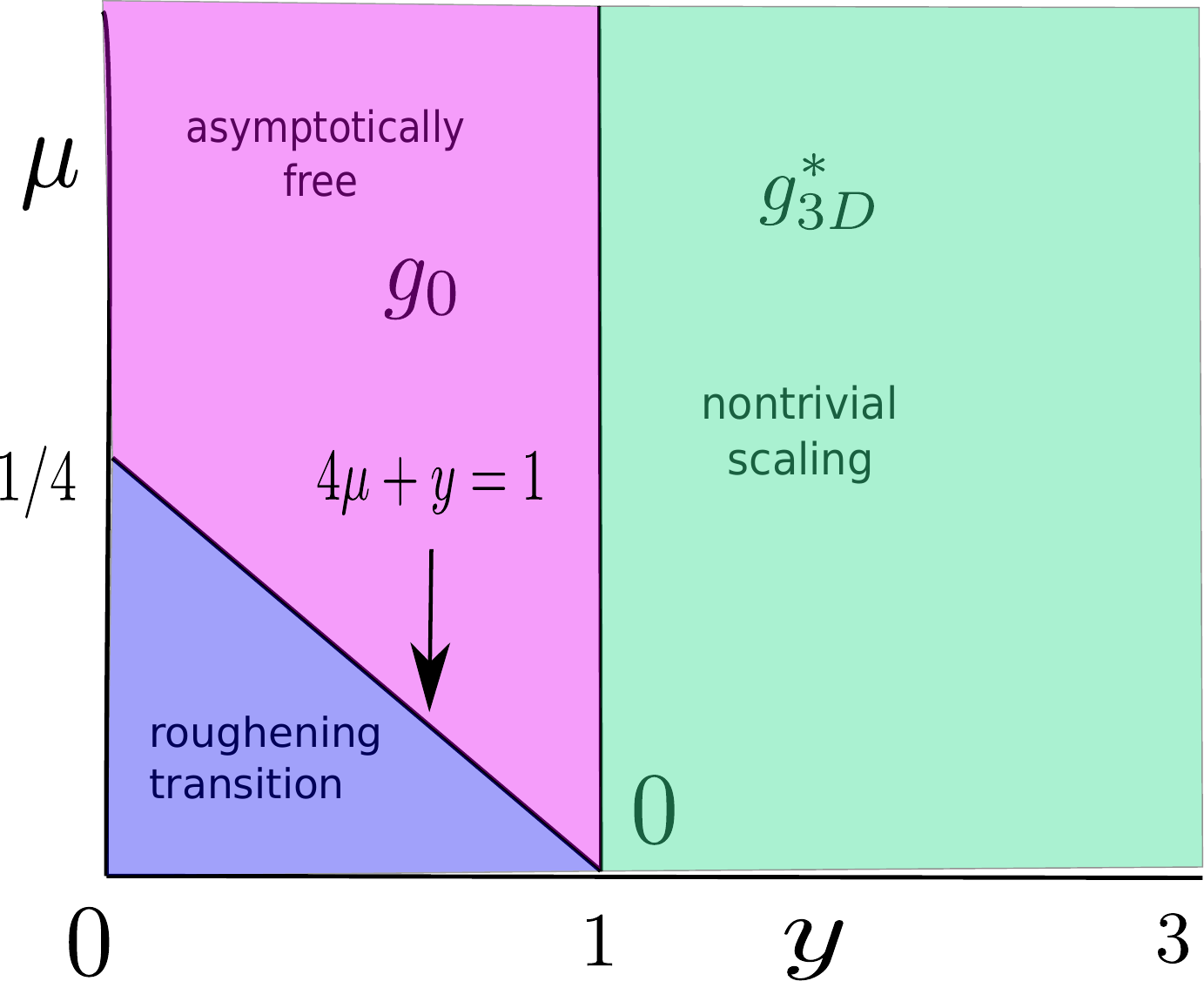}
 \caption{(a) Phase diagram of the 3D NRB XY model in the $y-\bar y$ plane. Phase LRO controlled by the nontrivial fixed point $g^*_{3D}$ (blue), phase SRO controlled by $g^*_{3D}$ (green), LRO controlled by $g_0=0$ (yellow) and SRO controlled by $g_0=0$ (red) are marked. The phase boundary between phase LRO and phase SRO, both controlled by $g_{3D}^*$ is given by $\chi=0$. (b) Phases in the $y-\mu$ plane: Phases controlled by $g^*_{3D}$ with nontrivial scaling different from those in the linear or free theory (phase LRO or phase SRO, green), roughening transition between asymptotically free phase LRO or SRO and perturbatively inaccessible strong coupling phases (violate) and asymptotically free with linear theory scaling controlled by the stable fixed point $g_0=0$  with no roughening transition (magenta). See text.}\label{phase-3d}
\end{figure}

\noindent
(ii)  In three dimensions (3D), relevant for the NRB XY model on a 3D lattice, the results are more dramatic. If $y>1$ {\em and} $4\mu+y-1>0$, then there is a stable phase, which can be  with orientational LRO if $[\bar y-1-(y-1)(4\mu-3+y)/(8\mu+2y-2)]/2<0$, or else positive with SRO, concomitant  with dynamics being faster or slower than ordinary diffusion. In contrast, if $y<1$ {\em and} $4\mu+y-1<0$, there is an unstable fixed point - a critical point - separating an asymptotically effectively noninteracting system with known scaling exponents, giving either LRO or SRO, and a perturbatively inaccessible strong coupling phase with unknown exponents. See Fig.~\ref{phase-3d}.

In both 2D and 3D, the scaling exponents vary continuously with $\mu$ for fixed $y,\bar y$.


We now derive our results by using the hydrodynamic equation for $\theta({\bf x},t)$, the local phase in the NRB XY model on a substrate and local surface height in the Monge gauge~\cite{nelson-book}, which reads~\cite{sm-ab}
\begin{eqnarray}
 \frac{\partial\theta}{\partial t} = \kappa{\nabla}^2\theta +\lambda{{\bf F}}\cdot {\boldsymbol\nabla}\theta + \bar\eta,\label{basic-hydro}
\end{eqnarray}
where $\kappa>0$ is the spin stiffness or surface tension. As shown in Ref.~\cite{sm-ab}, random quenched disordered vector $\bf F$ has its origin in the random nonreciprocal bonds in NRB XY and quenched advection for the surface, and $\lambda$ a coupling constant of arbitrary sign. In the absence of any conservation laws or other relevant dynamical variables, Eq.~(\ref{basic-hydro})  is the only hydrodynamic equation to consider. Equation~(\ref{basic-hydro}) a nonequilibrium or an ``active'' system, since the $\lambda$-term cannot be obtained from a free energy functional.
The random quenched disorder field $\bf F$ is a zero-mean, Gaussian-distributed with a variance in the Fourier space
\begin{eqnarray}
 \langle F_i({\bf k},\omega) F_j({\bf k}',\omega')\rangle &&= 2 \bigg[D_L Q_{ij}({\bf k}) + D_T P_{ij}({\bf k})\bigg]|k|^{-y}\nonumber \\ &&\times\delta ({\bf k+k}')\delta(\omega+\omega')\delta(\omega), \label{f-corr}
\end{eqnarray}
where $Q_{ij}({\bf k})=k_ik_j/k^2$ and $P_{ij}({\bf k})=\delta_{ij}-k_ik_j/k^2$ are respectively the longitudinal and transverse projection operators. Both $D_T$ and $D_L$ are positive, a requirement to ensure positivity of the variance of $\bf F(\bf k)$ at {\em all} $\bf k$~\cite{sm-ab}. The annealed noise $\eta({\bf x},t)$ is zero-mean, Gaussian-distributed with a variance in the Fourier space
\begin{equation}
 \langle\bar\eta({\bf k}),\omega)\bar\eta({\bf k}',\omega')\rangle = 2\bar D |k|^{-\bar y}\delta ({\bf k+k}')\delta (\omega+\omega').\label{noise-corr}
\end{equation}
With $y,\bar y>0$, both $\bf F({\bf k})$ and $\bar\eta ({\bf k},\omega)$ have {\em long-ranged} with spatially power-law or scale-free correlations, in contrast to Ref.~\cite{sm-ab}, where both of them are short-range. Correlations (\ref{f-corr}) and (\ref{noise-corr}) imply 
\begin{eqnarray}
 &&\langle F_i({\bf x})F_j({\bf x'})\rangle \propto |{\bf x-x'}|^{y-d},\\
 &&\langle \bar\eta ({\bf x},t)\bar\eta({\bf x'},t')\rangle \propto |{\bf x-x'}|^{\bar y-d},
\end{eqnarray}
in real space in $d$-dimensions, suppressing indices. Thus the disorder and noise correlations in real space are {\em non-local} and connect different distant parts of the system. We restrict $y,\bar y<d$ to ensure that the disorder and noise correlations {\em do not} grow with the separation $|{\bf x-x'}|$. 

 In the linearized approximation, i.e., with $\lambda=0$, for which the quenched disorder disappear, the variance of $\theta({\bf x},t)$ is given by
\begin{equation}
 \Delta_0\equiv \langle |\theta({\bf x},t)|^2\rangle_0 = \frac{\bar D}{\kappa}\int_{k=1/L}^{1/a_0} \frac{d^dk}{(2\pi)^d}|k|^{-\bar y -2},\label{linear-corr}
\end{equation}
where the subscript ``0'' refers to 
a linear theory result; $L$ and $a_0$ are the linear system size and a microscopic scale, respectively. From (\ref{linear-corr}), in 2D, $\Delta_0$ diverges with $L$ rapidly as $L^{\bar y}$ for any $\bar y>0$, giving only SRO.  In 3D, it is SRO for $\bar y>1$, whereas LRO is sustained by the linear theory for $0<\bar y<1$. 

Whether and to what extent the with quenched disorder can affect the linear theory results cannot be determined exactly due to the nonlinear nature of its coupling with $\theta$, for which the correlation function of $\theta$ cannot be calculated exactly. To go beyond the linear theory, we apply the perturbative dynamic Renormalization Group (hereafter ``RG'') to Eq.~\eqref{basic-hydro} at one-loop order~\cite{stanley,fns,uwe-book}. The fluctuation-induced corrections to the model parameters are captured by the corresponding one-loop Feynman graphs (see Appendix). There are diverging one-loop corrections to $\kappa, \lambda$, but not to $\bar D, D_L,D_T$. The absence of any relevant corrections to $\bar D$ is due to the fact that the noise variance (\ref{noise-corr}) is infra-red singular. The structure of the nonlinear vertex ensures that no corrections to the singular noise variance can be produced in a perturbative expansion. Dimensional analysis gives a dimensionless effective coupling constant $g$ and a dimensionless ratio $\mu$:
\begin{equation}
 g\equiv \frac{\lambda^2 D_L}{\kappa^2} k_d \Lambda^{d-2-y},\;\mu\equiv \frac{D_T}{D_L},\label{dim-ratio} 
\end{equation}
where $k_d$ is the surface area of a $d$-dimensional sphere of unit radius and $\Lambda\sim 1/a_0$ is an upper wavevector cutoff. Both $g$ and $\mu$ are positive definite. Since $D_L$ and $D_T$ are not renormalized, neither is $\mu$. We follow the standard RG procedure, analogous to that used in Refs.~\cite{medina,debayan-ckpz,sm-ab-mbe}, with the condition that the $\bf F$ is time-independent. Using a fixed dimension RG scheme~\cite{john-disorder3}, we evaluate the one-loop graphs at dimension $d$ and set $d=2,3$ to extract results at 2D and 3D, or alternatively, we can use a ``$\epsilon$-expansion about the critical dimension $d_c$ at which $g$ becomes marginal. Due to the structure of the model and the dependence of $d_c$ on $y$, both are closely related. The RG recursion relations for $\kappa,\lambda$ at the one-loop order are 
\begin{eqnarray}
 \frac{d\kappa}{d\ell}&=& \kappa\bigg[z-2 +g\bigg\{\mu\bigg(1-\frac{2}{d}\bigg) +\mu-1+\frac{y}{d}\bigg\}\bigg],\label{kappa-flow}\\
 \frac{d\lambda}{d\ell}&=& \lambda\bigg[z+\chi_F -1 -\frac{2g}{d}\bigg],\label{lambda-flow}
\end{eqnarray}
together with the exact recursion relation for $\bar D$
\begin{equation}
 \frac{d\bar D}{d\ell}= \bar D[z+\bar y -d -2\chi].\label{flow-Dbar}
\end{equation}
Here, $z$ and $\chi$ are the dynamic and spatial scaling exponents of $\theta({\bf x},t)$; $\chi_F$ is the spatial scaling exponent of ${\bf F}({\bf x})$. From (\ref{f-corr}), $\chi_F= (y-d)/2$. The RG flow equation for $g$ can be obtained by combining (\ref{kappa-flow})-(\ref{flow-Dbar}):
\begin{equation}
 \frac{dg}{d\ell}=g\bigg[y+2-d-g\bigg(\frac{4}{d}+4\mu -2-\frac{4\mu}{d} + \frac{2y}{d}\bigg)\bigg],\label{g-flow}
\end{equation}
with $d_c=y+2$ as the critical dimension. Solving (\ref{g-flow})
\begin{equation}
 g^*=\frac{y-d+2}{\frac{4}{d}+4\mu -2-\frac{4\mu}{d} + \frac{2y}{d}},\label{g-soln}
\end{equation}
as the nontrivial fixed point, in addition to $g_0=0$, the Gaussian fixed point. Since $g^*\geq 0$, we must have $y-d+2>(<)0$ and $\frac{4}{d}+4\mu -2-\frac{4\mu}{d} + \frac{2y}{d}>(<)0$ for $g^*$ being linearly stable (unstable). We now consider $d=2,3$ and focus on the stability of the fixed  points and the associated scaling exponents.

\noindent {(i)\em $d=2$:-} Setting $d=2$ in (\ref{g-soln}),
\begin{equation}
 g^*=\frac{y}{2\mu+y}\label{flowg-2d}
\end{equation}
is positive definite since $y>0$, $\mu\geq 0$, and is the only stable fixed point; $g_0=0$ is linearly unstable. In 2D, $d_c-d=y$. Hence, in an ``$\epsilon$-expansion about $d_c$~\cite{uwe-book}, $y$ appears as the small parameter, such that our results are asymptotically exact for $y\rightarrow 0$.  We now use (\ref{flowg-2d}) together with setting $d\kappa/d\ell =0,d\bar D/d\ell=0$ to get 
\begin{equation}
 z=2-\frac{y(2\mu-2+y)}{4\mu+2y},\,\chi=\frac{\bar y}{2}- \frac{y(2\mu-2+y)}{8\mu+4y},\label{expo-2d}
\end{equation}
which are parametrized by $\mu$ for given $y,\bar y$. For $ 2\mu-2+y<0$, $z>2$ and $\chi>0$ necessarily. Indeed, from (\ref{expo-2d}), $z\rightarrow 2$ as $y\rightarrow 0$. For sufficiently small $\mu$, $2\mu-2+y<0$. Hence, $z$ first increases from 2 as $y$ increases. This means slower than diffusive dynamics for both the 2D NRB XY model and the active surface, together with SRO in the NRB XY model. As $y\rightarrow 2-2\mu$, $z\rightarrow y$ again; for further rise in $y$, $z$ monotonically decreases as $y$ increases.  For a surface, if $\chi$ can be more or less than 1, corresponding to orientational SRO (presumably crumpled) or orientational LRO in the surface, depending upon the relative magnitudes of $y,\bar y$. These variations of $z$ and $\chi$ with $\mu$ are {\em continuous}. The condition $\chi=0$ gives
\begin{equation}
 \bar y = \frac{y(2\mu-2+y)}{4\mu+2y},\label{2dchi0}
\end{equation}
a surface in the $y-\bar y-\mu$ space on which $\chi=0$, giving QLRO phase order for the NRB XY model or positional QLRO and orientational LRO for the active surface. This surface is the phase boundary phase LRO and SRO for the NRB XY model, and positional LRO and SRO (but with orientaional LRO) in the $y-\bar y-\mu$ space. Similarly, the condition $\chi=1$ gives
\begin{equation}
 \bar y=2+\frac{y(2\mu-2+y)}{4\mu+2y},\label{2dchi1}
\end{equation}
a surface in the $y-\bar y-\mu$ space on which $\chi=1$, giving orientaional QLRO for the active surface, demarcating regions with orientaional LRO and SRO in the $y-\bar y-\mu$ space.

\noindent
{\em $d=3$:-} We now consider a bulk 3D NRB XY model. Then by using (\ref{g-soln}) and setting $d=3$
\begin{equation}
 g^*=\frac{3(y-1)}{8\mu + 2y -2} \label{flowg-3d}
\end{equation}
is the nontrivial fixed point, in addition to the Gaussian fixed point $g_0=0$. We note from (\ref{flowg-3d}) that $d=d_c=y-1$ plays the role of a small parameter in a potential ``$\epsilon$''-expansion in 3D.  Positivity of $g^*$ requires $y>(<)1$ and $8\mu + 2y -2>(<)0$. For $y>(<)1$, $g^*$ in (\ref{flowg-3d}) is a stable (unstable) fixed point.  When $g=g^*$ is an unstable (stable) fixed point, the Gaussian fixed is stable (unstable). The possible existence of an unstable nontrivial fixed point in some part of the parameter space make the 3D case fundamentally different and more complex than its counterpart in 2D. We now calculate $z$ and $\chi$ near $g=g^*$. From (\ref{kappa-flow}) and (\ref{flow-Dbar}) with $d=3$, we get  
\begin{eqnarray}
 &&z=2-  \frac{(y-1)}{8\mu + 2y -2}(4\mu+y-3),\label{z-3d}\\
 &&\chi=\frac{1}{2}\bigg[-1+\bar y - \frac{(y-1)(4\mu-3+y)}{8\mu + 2y -2}\bigg].\label{chi-3d}
\end{eqnarray}
Similar to the 2D results, $z$ can be more or less than 2 for $4\mu+y -3<(>)0$, giving slower (faster) than ordinary diffusive dynamics. Similarly, $\chi$ also can be negative or positive, depending upon the values of $y,\bar y,\mu$, giving LRO or SRO phase order. These values of the exponents hold regardless of $g^*$ is a stable or unstable fixed point. When $g^*$ is stable, these exponents give the scaling properties of phase fluctuations in the stable phase, which may or may not be phase-ordered. These exponents vary continuously with $\mu$. When $g^*$ is unstable, it implies a continuous transition at $g=g^*$, with the exponents in (\ref{z-3d}) and (\ref{chi-3d}) valid at the transition; it connects two stable phases, one at $g=g_0=0$ and another larger than $g^*$, which is a strong coupling phase that is inaccessible perturbatively. Interestingly, the phase controlled by the Gaussian fixed point is asymptotically ``free'' or noninteracting, i.e., $z=2$. However, $\chi$ can be more or less than 0, giving phase SRO or LRO.  We are not able to make any prediction on the scaling properties of the strong coupling phase, although we reasonably expect them vary with $\mu$. Independent of $\bar y$, for $(y,\mu)$ in specific sections of the $y-\mu$ plane, $g^*$ is stable or unstable. However, there are other regions in  the $y-\mu$ plane, in which neither the condition of a stable nor unstable $g^*$ are fulfilled. For example, for $\mu>(<)1$ and $ 4\mu-3+y<(>)0$, $g^*$ becomes negative, which is unphysical. In these cases, $g_0=0$ is the only stable fixed point.  See Fig.~\ref{phase-3d}.

Notice that in both 2D and 3D, by tuning $\mu$ phase transitions between states with LRO and SRO for fixed $y,\bar y$ can be induced in this model.



To summarize, we have studied the common hydrodynamic theory for a NRB XY model and an active surface advected by quenched tangential velocities, where the quenched disorder is assumed to be have a scale-free variance. Thus in the clean limit, the model only has SRO in 2D. The stochatic additive noise is also considered to be long-ranged, making this model an appropriate platform to study a fundamental issue of the consequence of the interplay and competion between corelated disorder and correlated noise. The surprising finding is that in this model is that since correlated quenched disorder can transport and redistribute phase fluctuations, it can increase the range of the quenched nonreciprocity can counteract the tendency of long-range annealed noise to destroy order. In particular, in both 2D and 3D, sufficiently long range disorder can suppress the order destroying nature of the noise and make the system to sustain LRO. In 3D, a competition between disorder and noise can also lead to a ``roughening transition'' between an asymptotically free state with LRO or SRO and a perturbatively inaccessible strong coupling phase of unknown exponents.  The scaling exponents in all the cases are nonuniversal, varying continuously with $\mu$ that controls the disorder distribution. This is analogous to Ref.~\cite{niladri-xy}  and Ref.~\cite{astik-qckpz}, where the scaling exponents vary continuously with a model parameter that control the noise and disorder distributions, respectively.

{\em Acknowledgment:-} S.M. and A.B. thank Alexander von Humboldt Stiftung (Germany)
for partial financial support through their research group linkage programme (2024). S.M. thanks ANRF (India) for partial financial
support through the ARG (MATRICS) programme (file no.: ANRF/ARGM/2025/000748/TS).  A.B. thanks ANRF (India) for partial financial
support through the ARG (MATRICS) programme (file no.:
ANRF/ARGM/2025/000461/TS).

\end{document}